\documentclass[12pt,a4paper,hypertex]{article}
\usepackage{jheppub}

\allowdisplaybreaks[1]
\usepackage{amsmath,bm}
\usepackage{subfigure}
\usepackage{color}
\usepackage{ulem}

\begin{document}

\title{
Confining solitons in the Higgs phase of ${\mathbb C}P^{N-1}$ model: Self-consistent exact solutions in large-$N$ limit
} 

\author[1,2]{Muneto Nitta} 
\author[2,3]{and Ryosuke Yoshii} 
\affiliation[1]{Department of Physics, Keio University, 4-1-1 Hiyoshi, Kanagawa 223-8521, Japan}
\affiliation[2]{Research and Education Center for Natural Sciences, Keio University, 4-1-1 Hiyoshi, Kanagawa 223-8521, Japan}
\affiliation[3]{Department of Physics, Chuo University, 1-13-27 Kasuga, Bunkyo-ku, Tokyo 112-8551, Japan}
\emailAdd{nitta@phys-h.keio.ac.jp}
\emailAdd{yoshii@phys.chuo-u.ac.jp}

\date{\today}
\abstract{
The quantum ${\mathbb C}P^{N-1}$ model is in the confining (or unbroken) phase with a full mass gap in an infinite space, 
while it is in the Higgs (broken or deconfinement) phase accompanied with Nambu-Goldstone modes in a finite space such as a ring or finite interval smaller than a certain critical size.
We find a new self-consistent exact solution describing a soliton in the Higgs phase of the ${\mathbb C}P^{N-1}$ model in the large-$N$ limit on a ring. 
We call it a confining soliton.
We show that all eigenmodes have real and positive energy and thus it is stable.
}

\maketitle

\section{Introduction}

The two-dimensional ${\mathbb C}P^{N-1}$ model \cite{Eichenherr:1978qa, Golo:1978dd, Cremmer:1978bh} 
shares many common non-perturbative phenomena 
with four dimensional Yang-Mills theory, such as
confinement, asymptotic freedom, dynamical mass generation, 
and instantons \cite{Polyakov:1975rr, Polyakov:1975yp, Bardeen:1976zh, Brezin:1976qa, DAdda:1978vbw, DAdda:1978etr, Witten:1978bc,Novikov:1984ac}. 
The two-dimensional Gross-Neveu (GN)  \cite{Gross:1974jv} or 
Nambu Jona-Lasino (NJL) model \cite{Nambu:1961tp} exhibits 
dynamical mass gap and dynamical chiral symmetry breaking, 
with sharing these properties with four dimensional QCD. 
These two-dimensional theories, one consisting of bosons with nonlinear interaction and the other describing fermions with a four Fermi interaction,  
have been well studied in the large-$N$ or mean field approximation  
(where $N$ is the number of flavors for the GN model), 
as toy models to study non-perturbative effects in four-dimensional gauge theories. 
 
The ${\mathbb C}P^{N-1}$ model is in the confining (or unbroken) phase with a mass gap when it is defined in an infinite space, 
in agreement with the Coleman-Mermin-Wagner (CMW) theorem 
prohibiting a spontaneous symmetry breaking or a long-range order in 1+1D systems  \cite{Coleman:1973ci, Mermin:1966fe}. 
While this can be well shown in the large-$N$ or mean field approximation in general,  
the mass gap can be exactly shown by using the mirror symmetry \cite{Hori:2000kt} in the case of the supersymmetric ${\mathbb C}P^{N-1}$ model \cite{Witten:1977xn, DiVecchia:1977nxl}.  
On the other hand, the Higgs (broken or deconfinement) phase 
accompanied with Nambu-Goldstone (NG) modes 
is possible when the ${\mathbb C}P^{N-1}$ is defined in a finite system; 
Recently phase transitions between the confining and Higgs phases in finite systems have been discovered in the ${\mathbb C}P^{N-1}$ model on a ring \cite{Monin:2015xwa, Monin:2016vah} and on a finite interval \cite{Milekhin:2012ca, Milekhin:2016fai, Pavshinkin:2017kwz}. 
In these cases, the Higgs phase is favored when the system size is smaller than some critical size. 
It is consistent with the CMW theorem because, in a finite size, this phase does not have a long range order correlation inhibited by the theorem. 

The GN model is in the broken phase in the large-$N$ or mean field approximation, in which the order parameter is constant and the ${\mathbb Z}_2$ symmetry 
is dynamically broken.
The chiral GN (or NJL) model, having a complex order parameter, is in the broken phase (or Bardeen-Cooper-Schrieffer (BCS) phase in the context of 
superconductivity) in the ground state in the infinite system in the large-$N$ limit. 
Though one might think this phase to be incompatible with the CMW theorem because of $U(1)$ symmetry breaking, 
it was shown to be compatible with the CMW theorem in 
Ref.\ \cite{Witten:1978qu}.

The above arguments only consider homogeneous configurations.  
Although it is usually the case that a constant condensation is the lowest energy state in a homogeneous system without any external field, 
inhomogeneous condensations may appear as the lowest energy states in certain situations. 
For instance, a twisted boundary condition unavoidably make a configuration inhomogeneous, {\it e.g.}, 
the 1D Ising model with the anti-parallel spin orientation on the opposite boundary. 
Many examples of inhomogeneous configurations have been known for long time 
in the (chiral) GN model, 
corresponding to conducting polymers or superconductors in physical systems. 
For instance, a phase twist 
on a superconductor results in the Fulde-Ferrell (FF) state  \cite{Fulde:1964zz}, in which the order parameter is plane-wave like.
Another possibility is the presence of the external fields;  
In the case of superconductors, an applied magnetic field induces the spin excess by the Zeeman effect and thus the order parameter becomes nodal \cite{Machida:1984zz}, which is known as the Larkin-Ovchinnikov (LO) state \cite{Larkin:1964zz} or a real kink crystal. A self-consistent exact solution for it was obtained in Ref.~\cite{Brazovskii}. 
Those self-consistent solutions were generalized to a system with a spin imbalance \cite{Yoshii:2011yt, Takahashi:2012aw}. 
A self-consistent analytic solution of the FFLO state or a twisted kink crystal was found in Ref.~\cite{Basar:2008im}, 
in which both the amplitude and phase of the order parameter modulate. 
A superconducting ring with a Zeeman magnetic field and magnetic flux penetrating the ring exhibits similar configurations \cite{Yoshii:2014fwa}. 
Inhomogeneous configurations also appear in a finite interval with the Dirichlet boundary condition \cite{Flachi:2017cdo}. 
Apart from inhomogeneous ground states, solitons as excited states have been constructed as self-consistent analytic solutions, such as a kink \cite{Dashen:1975xh, Takayama:1980zz}, 
kink-anti-kink \cite{Dashen:1975xh, Campbell:1981dc}, kink-anti-kink-kink \cite{Okuno:1983, Feinberg:1996gz, Feinberg:2002nq} for the real GN model, and twisted kink \cite{Shei:1976mn} and multiple twisted kinks with arbitrary separations \cite{Takahashi:2012pk} in the chiral GN model. 

On the other hand, 
inhomogeneous or soliton solutions for the {\it quantum} 
${\mathbb C}P^{N-1}$ model had not been discussed until recently, 
while various {\it classical} solitons in the ${\mathbb C}P^{N-1}$ model have been known for long time, such as kinks (domain walls) \cite{Abraham:1992vb,Gauntlett:2000ib,Isozumi:2004jc}, lumps (instantons) \cite{Polyakov:1975yp}, and their composites \cite{Gauntlett:2000de,Isozumi:2004vg}.
In the case of a finite interval with  the Dirichlet boundary condition,
it is inevitable that the Higgs field and mass gap function are inhomogeneous \cite{Bolognesi:2016zjp, Betti:2017zcm,Flachi:2017xat,Nitta:2018lnn,Bolognesi:2018njt}. 
Self-consistent analytic solutions for this case was obtained in Refs.~\cite{Flachi:2017xat,Nitta:2018lnn}. 
A similar problem in a two-dimensional disk was discussed in Refs.~\cite{Gorsky:2013rpa,Pikalov:2017lrb}.

\begin{table}
\begin{center}
\includegraphics[width=15pc]{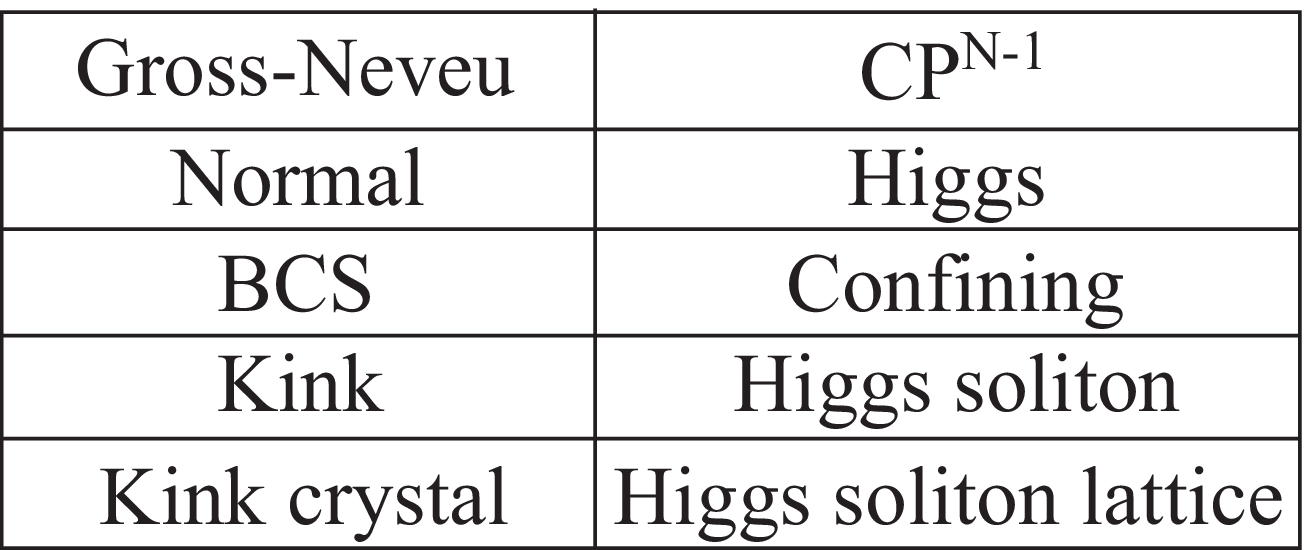}
\caption{The correspondence between the previously known solutions in the GN model and those in the ${\mathbb C}P^{N-1}$ model.} 
\label{TableCPN-GN}
\end{center}
\end{table} 
The hidden connection between the ${\mathbb C}P^{N-1}$ and GN models has been recently revealed; 
the self-consistent equations obtained in the leading order of the large-$N$ limit for the both theories coincide \cite{Nitta:2017uog}.
This correspondence provides a map from solutions in the GN model to those in the ${\mathbb C}P^{N-1}$ model, and by using it, 
we have constructed several self-consistent inhomogeneous solutions of the ${\mathbb C}P^{N-1}$ model as summarized in Table \ref{TableCPN-GN}. 
One is a self-consistent soliton solution in the confining phase of the ${\mathbb C}P^{N-1}$ model in an infinite space or on a ring \cite{Nitta:2017uog}. 
This soliton is stable in the linear order because all eigenmodes have real and non-negative energy.  
The Higgs phase with NG modes is localized inside the soliton, thereby called the Higgs soliton. 
Multiple Higgs solitons with arbitrary separation have also been obtained. The twisted boundary condition results in the flavor rotating solutions in 
the ${\mathbb C}P^{N-1}$ model \cite{Nitta:2018lnn} 
(see also \cite{Bolognesi:2018njt}).

A natural question arising immediately is 
whether any soliton can exist in the Higgs phase of the ${\mathbb C}P^{N-1}$ model.
In this paper, we construct self-consistent analytical soliton solutions of a kink-shape or kink-anti kink crystal shape in the Higgs phase of the ${\mathbb C}P^{N-1}$ model.
The solution is LO-like in the sense that the amplitude of the Higgs field modulates. 
We call them confining solitons since the confining phase is localized inside the solitons.
Confining solitons are stable at least in the linear order; 
all eigenmodes are real and non-negative. 
While {\it classical} solutions such as kinks (domain walls),  
lumps (instantons) and their composites were known in the Higgs phase in the ${\mathbb C}P^{N-1}$ model, 
our new solitons in this paper are the first examples of {\it quantum} solitons in the Higgs phase of the ${\mathbb C}P^{N-1}$ model. 

Since the 1+1D ${\mathbb C}P^{N-1}$ model appears on a non-Abelian vortex string in 3+1D (supersymmetric) $U(N)$ gauge theory 
\cite{Hanany:2003hp, Auzzi:2003fs, Eto:2005yh, Tong:2005un, Eto:2006pg, Shifman:2007ce, Tong:2008qd, Hanany:2004ea, Shifman:2004dr} 
or dense QCD (for $N=3$) \cite{Nakano:2007dr,Eto:2013hoa}, 
the above studies of quantum phases of the ${\mathbb C}P^{N-1}$ model describe quantum vortex states. 
The ${\mathbb C}P^{N-1}$ model on a ring and interval describe 
a closed string and open string stretched between monopoles or domain walls, respectively.

The $O(3)$ sigma model which is equivalent to the ${\mathbb C}P^{1}$ model 
also appears in the condensed matter physics. 
It is the continuous model of the Heisenberg anti-ferromagnetic spin chain \cite{Haldane:1982rj, Affleck}, and the quantum phase transition, so-called deconfined criticality, is proposed in the anti-ferromagnetic system \cite{Senthil, Nogueira:2013oza}. 
Our solution may be relevant to this case.

This paper is organized as follows. 
In Sec.\ \ref{MM}, we briefly review our method to obtain self-consistent solutions. 
In Sec.\ \ref{CSS}, a kink solution is introduced and the existence of the solution in  finite systems is discussed. 
In Sec.\ \ref{CSCS}, the solution in Sec.\ \ref{CSS} is generalized to a kink crystal solution. 
Finally we summarize the results of our paper and discuss future problems in Sec.\ \ref{Summary}.

\section{Model and method}\label{MM}
The action of the  ${\mathbb C}P^{N-1}$ model is given by
\begin{equation}
S=\int dtdx \left[(D_\mu n_i)^\ast (D^\mu n_i)-\lambda (n_i^\ast n_i-r)\right], 
\label{action}
\end{equation}
where the first term describes the kinetic term of $N$ complex scalar fields $n^i$ ($i=1,\cdots, N$) with the covariant derivative $D_\mu=\partial_\mu-iA_\mu$, 
and the second term gives the constraint $n_i^\ast n_i=r$ by the Lagrange multiplier $\lambda(x)$. 

When this model is considered as the effective theory of a non-Abelian vortex in $U(N)$ gauge theory, the ``radius" $r$ can be written as $r=4\pi/g_{\rm YM}^2$, where $g_{\rm YM}$ is a coupling constant in the Yang-Mills theory. 

Here we note that the gauge field in this model is auxiliary (no kinetic term) 
and thus we set $A_\mu=0$ in the following. 
One can obtain the effective action for $n_1\equiv \sigma$ by integrating out the rests $n_i$ ($i=2, \cdots, N$): 
\begin{align}
S_{\rm eff}=\int dt dx \left[(N-1) \mathrm{Tr} \ln (-\partial_\mu \partial^\mu +\lambda)
+\partial_\mu\sigma\partial^\mu\sigma-\lambda (\sigma^2-r)\right]. 
\label{effaction}
\end{align}
The total energy functional can be written as  
\begin{equation}
E=(N-1)\sum_n\omega_n+\int dx \left[(\partial_x\sigma)^2+\lambda(\sigma^2-r)\right]. \label{eq:energy0}
\end{equation}
The stationary conditions for $\lambda$ and $\sigma$, respectively, yield \cite{Bolognesi:2016zjp} 
\begin{align}
&\frac{N-1}{2}\sum_n \frac{|f_n|^2}{\omega_n}+\sigma^2-r=0, \label{gapeq}\\
&(-\partial_x^2+\lambda)\sigma=0, \label{eqsigma}
\end{align}
where $f_n(x)$ and $\omega_n$ are given as orthonormal eigenstates and eigenvalues of the following equation 
\begin{equation}
(-\partial_x^2 +\lambda)f_n(x)=\omega_n^2 f_n(x). \label{eigeneq}
\end{equation}
Thus we end up with three Eqs.~(\ref{gapeq})--(\ref{eigeneq}). 
We here note that Eqs.~(\ref{eqsigma}) is nothing but the zero mode equation of Eq.\ (\ref{eigeneq}). 

In the previous paper \cite{Nitta:2017uog}, we have found a map from above self-consistent equations to the gap equation and eigenvalue equation for the GN model. 
We have introduced the auxiliary field $\Delta$ as 
\begin{equation}
\Delta^2+\partial_x\Delta=\lambda(x), 
\label{deflambda}
\end{equation}
and find the solution for Eq.\ (\ref{eqsigma}), 
\begin{align}
\sigma=A\exp \left[\int^x dy \Delta(y)\right], \label{sigmasol}
\end{align}
with is the integral constant $A$ which must be determined from Eq.\ (\ref{gapeq}). 

The energy functional Eq.~(\ref{eq:energy0}) becomes 
\begin{equation}
E_{\mathrm{tot}}=N\sum_n\omega_n-r\int^\infty_{-\infty} dx \lambda(x)
+ \left.\sigma\partial_x \sigma\right|_{-\infty}^{\infty}.\label{eq:energy}
\end{equation}
The important point is that the real solution $\Delta$ (order parameter) for the GN model automatically satisfies the 
self-consistent equations in the $\mathbb{C}P^{N-1}$ model. 
Thanks to this map we have found the various new solutions which are inhomogeneous in the space. 
In the present paper, we find a new solution with kink crystal Higgs profiles and also discuss a possible setup in which the new solution can exist.

\section{Confining soliton in the Higgs phase}\label{CSS}
As we already mentioned, the homogeneous confining and Higgs phases were known in the ${\mathbb C}P^{N-1}$ model.  
Those solutions are mapped from $\Delta=\mathrm{const}$ (BCS phase) and $\Delta=0$ (normal phase), respectively
in the GN model. 
In addition, we have found the following inhomogeneous solutions (the corresponding solution in the GN model are denoted in the brackets in the following, see also Table\ \ref{TableCPN-GN}); 
localized Higgs soliton (kink solution) and the localized Higgs soliton lattice solution (kink crystal solution),  
two Higgs soliton solution with the arbitrary separation (kink-anti-kink-kink solution), and the ``confining" solution (BCS-like solution) and the ``Higgs" solution (normal-like solution) in a finite interval. 

We start from a brief review for the homogeneous Higgs phase in a periodic system with the length $L$.  
This phase corresponds to $\Delta=0$ and thus 
\begin{equation}
\lambda=0,\ \sigma=A.
\label{Higgs}
\end{equation} 
In this case, the eigenfunctions and eigenvalues are, respectively, given as 
\begin{equation}
f_k(x)=\sqrt{\frac{1}{L}}e^{ikx},\ k=\frac{2\pi n}{L}, 
\label{EigenHiggs}
\end{equation}
where $n$ is an integer. 
By substituting Eqs.\ (\ref{Higgs}) and (\ref{EigenHiggs}) into the gap equation (\ref{gapeq}), 
we obtain
\begin{align}
\frac{N-1}{2L} \sum_{n=1} \frac{L}{2\pi n}+A^2-r=0. 
\end{align}
This equation allows the normalized zero mode $A=\sqrt{1/L}$ with renormalized coupling constant $r$. 
After the continuous approximation, the situation becomes more clearer. 
By using the replacement $2\pi/L\sum_n=\int dk$, one obtains 
\begin{align}
\frac{N-1}{4\pi} \int_{2\pi/L}^{\Lambda_{UV}} dk \frac{1}{k}+A^2-r
=\frac{N-1}{4\pi} \ln \frac{\Lambda_{UV}}{2\pi/L} +A^2-r=0. 
\end{align}
This shows that the size $L$ introduces a low-energy cutoff and the solution is forbidden in the infinite size limit $L\to \infty$ since $r$ must diverges as $\ln (2\pi/L)$ in the limit.  

\begin{figure}
\begin{center}
\includegraphics[width=20pc]{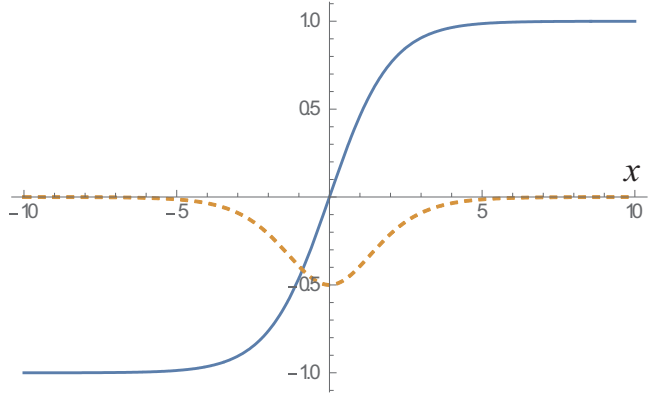}
\caption{The configuration of $\sigma$ (solid line) and $\lambda$ (dashed line)
Here we normalize as peak of $\sigma$ to be $1$ and $m=1$. }
\label{FigKink}
\end{center}
\end{figure}

Now we move to inhomogeneous solutions.
A candidate of a new solution would be the following
\begin{equation}
\sigma=A\tanh \frac{mx}{2},\quad \lambda=-\frac{m^2}{2\cosh^2 mx/2}.
\label{Higgskinksols}
\end{equation}
In Fig.\ \ref{FigKink}, we plot the Higgs field configuration and the corresponding mass gap function of the configuration in Eq.~(\ref{Higgskinksols}).
The Higgs field $\sigma$ exhibits a kink profile for this solution, 
whereas the mass gap function has a dip at the node of the kink. 
This configuration is obtained through the map in the last section from 
a singular solution 
\begin{equation}
 \Delta=m/\sinh mx \label{eq:GN}
\end{equation} 
in the GN model.
While the singular point is present at $x=0$ in Eq.~(\ref{eq:GN})
in the GN model, 
the configuration in Eq.~(\ref{Higgskinksols}) in the ${\mathbb C}P^{N-1}$ model is regular everywhere.
When we restrict Eq.~(\ref{eq:GN}) to the $x>0$ region, 
it is a solution on a half infinite line with the Dirichlet boundary condition, which reduces to the normal phase in the limit of the boundary sent to the infinity $x=-\infty$. 

Though the mass gap function in Eq.~(\ref{Higgskinksols}) takes negative values everywhere and seems to be problematic, unstable modes are absent. 
In fact, the above mass gap function is called the P\"oschl-Teller potential and all the energy spectra for this potential are known to be positive \cite{PoschlTeller, Lekner}. 
The eigenfunctions are explicitly given as 
\begin{equation}
f_k= \frac{1}{\sqrt{2}}\frac{4}{\sqrt{k^2+m^2}}e^{ikmx/2} (ik-m\tanh mx/2),
\label{HiggskinkEigenfns}
\end{equation}
where the eigenvalues are given as 
\begin{equation}
\omega_k=k/2,  \quad k>0.
\end{equation}
A set of configurations in Eqs.~(\ref{Higgskinksols}) and (\ref{HiggskinkEigenfns}) satisfies Eqs.\ (\ref{eqsigma}) and (\ref{eigeneq}). 

For the map between the ${\mathbb C}P^{N-1}$ and GN models, 
we use Eq.\ (\ref{gapeq}) differentiated with respect to $x$ instead of Eq.\ (\ref{gapeq}) itself. 
Thus the left hand side of Eq.\ (\ref{gapeq}) can be shown to be constant, but the map of the solution in the GN model does not necessary yield a  solution of the $\mathbb{C}P^{N-1}$ model.\footnote{ 
In fact the homogeneous Higgs solution is known to be inappropriate as the solution in the infinite system, 
though the corresponding solution so-called normal phase exists in the GN model.}
For the present case, the first term in the left hand side of the gap Eq.\ (\ref{gapeq}) becomes 
\begin{equation}
\text{const} \times \int^{\Lambda_{\mathrm{UV}}}_{\Lambda_{\mathrm{IR}}} 
\frac{1}{k}\frac{k^2 +m^2\tanh^2 mx/2}{k^2+m^2}, 
\end{equation}
where $\Lambda_{\mathrm{UV}}$ and $\Lambda_{\mathrm{IR}}$ are the ultraviolet and infrared cutoff, respectively. 
In the case of the infinite system, 
$\Lambda_{\mathrm{IR}}\rightarrow 0$ and the gap equation cannot be satisfied, which is consistent with the CMW theorem. 
However if we consider a finite system with the length $L$ the infrared cutoff $\simeq 1/L$ because the quantization of the energy levels are introduced, and thus a solution can exist. 
For instance, a ring of the length $L$ with anti-periodic boundary condition in which the Higgs field changes the sign 
($\pi$ rotation) $\sigma(0)=-\sigma(L)$ allows a solution, as discussed in the next section.

For this configuration all eigenmodes have real and non-negative energy, and thus the solution is energetically stable in the linear order. 
The minimum energy state is homogeneous solution as follows. 
The energy difference between the homogeneous Higgs solution with the confining soliton and the homogeneous Higgs phase can be calculated as 
\begin{equation}
E_{\mathrm{kink}}-E_{\mathrm{Higgs}}
=-r\int dx \left(-\frac{m^2}{2\cosh^2 mx/2}\right)
=rm^2.
\end{equation}
Here we have used the fact that the energy spectra for both solutions are the same and thus the energy difference only comes from the second term in Eq.\ (\ref{eq:energy}).

Here we note a similarity and difference from the Higgs soliton solution 
in the confining phase, 
which we obtained before. 
In the Higgs soliton solution \cite{Nitta:2017uog} 
\begin{equation}
\sigma=\frac{A}{\cosh mx},\ \quad
\lambda=m^2\left(1-2\frac{1}{\cosh^2 mx}\right),
\end{equation}
the localized Higgs field appears in the confining background. 
The corresponding mass gap function has a dip at the peak of the Higgs field 
in the background constant mass gap. 
The energy spectra  and the mass gap parameter $m$ coincide with those of the homogeneous confining phase. 

On the other hand, in the present solution, 
the localized confining part ($\sigma=0$) or localized confining soliton appears in the Higgs phase.
Thus we call this solution as a confining soliton. 
The mass gap function has a dip around the confining soliton and it becomes zero far away from the soliton. 
The energy spectra of the confining soliton solution coincide with those of the homogeneous Higgs phase. 
The confining soliton solution contains the mass parameter $m$, 
although the homogeneous Higgs phase does not have it.

It may be interesting to note that this solution is similar to the kink solution in the GN model \cite{Dashen:1975xh,Takayama:1980zz}; 
the order parameter for the kink solution is also given by the hyperbolic tangent $\Delta=M\tanh Mx$.  
However, this coincidence is no longer valid in the soliton lattice case as we will see in the next section.

\section{Confining soliton lattice solution} \label{CSCS}
\begin{figure}
\begin{center}
\includegraphics[width=20pc]{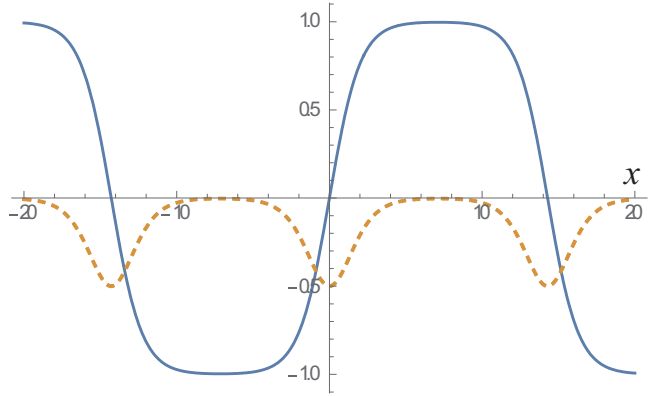}
\caption{The localized confining soliton lattice configuration of $\sigma$ (solid line) and $\lambda$ (dashed line) for  $\nu=1-10^{-5}$. 
Here we set, $m=1$ and normalize the peak of $\sigma$ to be $1$. }
\label{FigKinkCrystal}
\end{center}
\end{figure}

\begin{figure}
\begin{center}
\includegraphics[width=35pc]{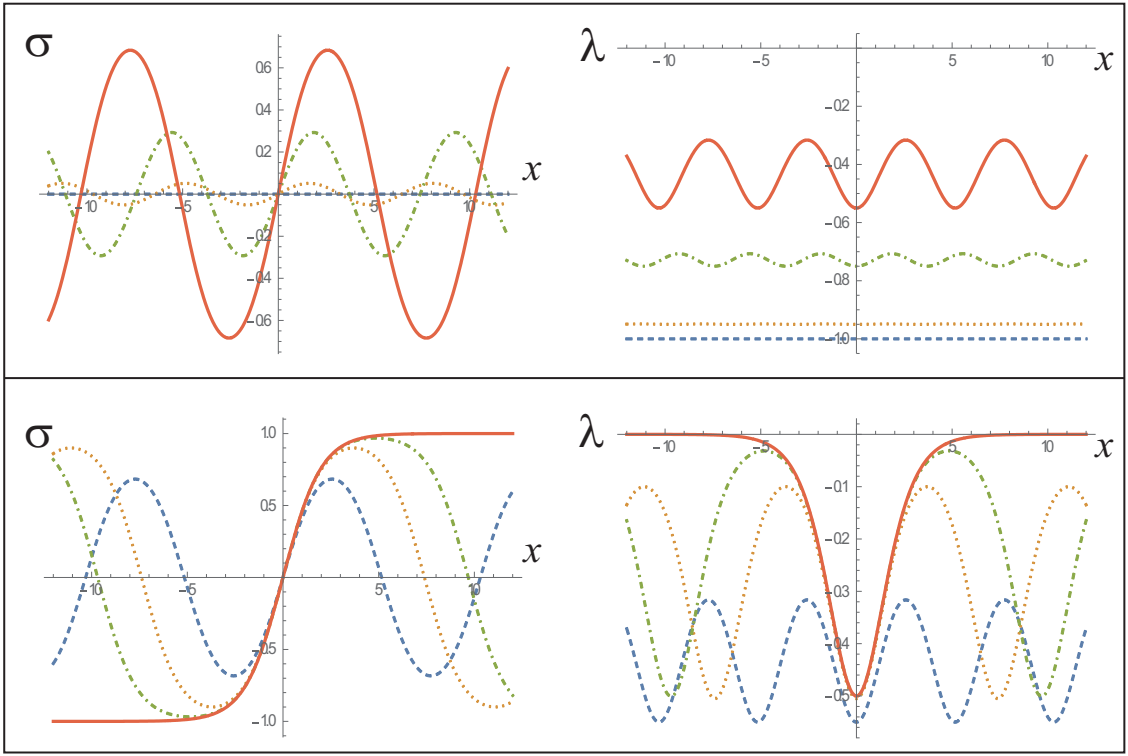}
\caption{The localized confining soliton lattice configuration with different $\nu$. 
The upper figures correspond to $\nu=0$ (dashed line), $\nu=0.1$ (dotted line), $\nu=0.5$ (dot-dashed line), and $\nu=0.9$ (solid line).  
The lower figures correspond to $\nu=0.9$ (dashed line), $\nu=0.99$ (dotted line), $\nu=0.999$ (dot-dashed line), and $\nu=1$ (solid line). 
Here we set, $m=1$ and normalize the peak of $\sigma$ to be $1$. }
\label{FigDiffnu}
\end{center}
\end{figure}

In the last section, we have considered the single confining soliton.
This configuration is a special case of the following solution 
\begin{equation}
\sigma=A_{CSL}\frac{-\mathrm{dn}(mx,\nu)+1}{\mathrm{sn}(mx,\nu)},\ \quad
\lambda=\frac{\mathrm{cn}^2(mx,\nu)-\mathrm{dn}(mx,\nu)}{\mathrm{sn}^2(mx,\nu)},
\label{CSC}
\end{equation}
where $\mathrm{sn},\ \mathrm{cn},\ \mathrm{dn}$ are Jacobi's elliptic functions, $\nu$ is the elliptic parameter, 
and $A_{CSL}$ is a normalization constant. 
Here the $A_{CSL}$ is determined by the gap equation (\ref{gapeq}) as 
\begin{equation}
A_{CSL}^2={s(x)^{-2}}\left(r-\frac{N-1}{2}\sum_n \frac{|f_n^{CSL}|^2}{\omega_n}\right),
\label{gapeqexpl}
\end{equation}
where $s(x)=\sigma(x)/A_{CSL}$ and $f_n^{CSL}$ is the eigenfunctions obtained from Eq.\ (\ref{eqsigma}) for the current solution. 
The mapping between the GN and ${\mathbb C}P^{N-1}$ models ensures that the left hand side of Eq.\ (\ref{gapeqexpl}) becomes a constant.
Though it can be fixed by a numerical summation, it is difficult to give an explicit form of the normalization factor for the present solution.
In Fig.\ \ref{FigKinkCrystal}, we plot the Higgs fields and the mass gap functions for various parameters.
In this configuration, the Higgs field exhibits a kink crystal structure 
and the mass gap function does a soliton lattice structure. 
This configuration describes a confining soliton lattice in the Higgs phase. 
The periodicity $l$ of the solution is given by the complete elliptic integral of the first kind, 
\begin{equation} 
l =2\mathrm{K}(\nu).
\end{equation} 
The single soliton configuration presented in the last section is corresponding to the case of $\nu=1$. 
For this solution, we need to be careful since the energy spectra are not always positive in the infinite system. 
For instance, if we consider the $\nu=0$ limit, the mass gap function becomes 
$\lambda=-m^2$ (Fig.\ \ref{FigDiffnu}). 
This means that the energy spectra start from $-m^2$ 
which does not cause unstable modes when the size of the system is smaller than $L_{C}= \pi/m$, 
since the system with the length $L$ has the lowest kinetic energy $\sim \pi^2/L^2$. This means that the system with the size $L$ cannot have $m$ larger than $\sim \pi/L$. 
The condition $L\ll 1/m$ for the existence of the stable solution is the same with one for the existence of the homogeneous Higgs solution in the cases of a ring \cite{Monin:2015xwa,Monin:2016vah} and a finite interval \cite{Milekhin:2012ca, Milekhin:2016fai, Pavshinkin:2017kwz}.

Let us make a comment on the similarity between solutions in the ${\mathbb C}P^{N-1}$ and GN models. 
The solution has an amplitude modulation of the Higgs field and thus it can be considered as a counterpart of the LO solution in the GN model, 
although the solution for the soliton lattice solution is not identical to the LO solution $\Delta=M \mathrm{sn} (Mx, \nu)$ \cite{Brazovskii} in the GN model.  As we saw in the last section, only in the single kink limit, those functions coincide.

Using the configuration in Eq.\ (\ref{CSC}), we can construct a self-consistent analytic solutions on a ring. 
Here we consider (a) periodic or (b) anti-periodic boundary conditions. 

(a)
For the case of the periodic boundary condition, one can satisfy the condition by choosing $\nu$ such that the $nl$ ($n \in {\mathbb Z}$) to be the size of the ring. 
In this case, the homogeneous Higgs solution also satisfies the same boundary condition. 

(b)
For the case of the anti-periodic boundary condition, one needs to choose $\nu$ to satisfy $L=(n+1/2)l$ with $n \in {\mathbb Z}$. 
In this case the homogeneous Higgs solution with the flavor rotation \cite{Nitta:2018lnn} can also be the solution of the same boundary condition.

\section{Summary}\label{Summary}
In this paper, we have constructed the self-consistent localized confining soliton solution in the Higgs phase of the ${\mathbb C}P^{N-1}$ model.
While this solution is inhibited in the infinite system as the homogeneous Higgs phase is inhibited in the same limit, it is allowed in finite systems due to the existence of the infrared cutoff  $\sim 1/L$. 
Our solutions are stable since the all eigenmodes have real and positive energy. 
We have generalized the kink solution to the case of the soliton lattice. 

Our present calculation relies on the large-$N$ limit, 
however the resulting self-consistent equations are 
the same with the equations obtained by the mean field approximation. 
Thus we expect that our solution is qualitatively reliable even for the case of $N\sim 1$. 

Throughout the present paper, we have discussed the $\mathbb{C}P^{N-1}$ model, but the same analysis is applicable to the $O(N)$ model. 
Thus the present solutions are expected to appear in the spin system.  

Although our solutions are stable at linear order, 
the minimum energy state should be the homogeneous Higgs phase.
It remains an interesting future problem to look for conditions that the confining soliton lattice becomes the minimum energy state. 
For instance, if we consider the anti-periodic boundary condition, solutions satisfying the same boundary condition known so far are 
the confining soliton in this paper and the flavor rotating solutions obtained in the previous paper. 
Suppose we have an anisotropy which makes Higgs field tend to align in a specific direction, 
we expect that the confining kink solution is favored since the flavor rotating solution costs more energy to have the components not aligned to the selected direction. 
Another possibility could be the presence of a defect. 
For example, if the Higgs fields are obliged to vanish at the defect, the kink solution might be chosen energetically. 
We leave this for the future problem. 

Our solution may be relevant for a gauge field localization in the brane world scenario, see, e.g., Refs.\ \cite{Dvali:1996xe, Ohta:2010fu}. 
This is because $U(1)$ gauge field is absent outside the soliton in the Higgs phase while it is present as a composite gauge boson inside the soliton. In order to make this more realistic we have to introduce more dimensions orthogonal to the soliton to yield it the world-volume directions.

The inhomogeneous solutions obtained so far are given by the ${\mathbb C}P^{N-1}$/GN correspondence, 
in which we can use only real condensates of the real GN model. 
Thus, a possible generalization should be the case of the chiral GN model which has the complex order parameters. 
We also leave this for the future problem.

\section*{Acknowledgement}
The support of the Ministry of Education,
Culture, Sports, Science (MEXT)-Supported Program for the Strategic Research Foundation at Private Universities `Topological Science' (Grant No.\ S1511006) is gratefully acknowledged. 
The work of M.~N.~is 
supported in part by the Japan Society for the Promotion of Science
(JSPS) Grant-in-Aid for Scientific Research (KAKENHI Grant
No.~16H03984 and 18H01217) 
and by a Grant-in-Aid for Scientific Research on Innovative Areas ``Topological Materials
Science'' (KAKENHI Grant No.~15H05855) from the MEXT of Japan.

\end{document}